\documentclass[%
aps,amsmath,amssymb, superscriptaddress,
reprint,prb%
]{revtex4-2}
\usepackage[normalem]{ulem}
\usepackage{graphicx}
\usepackage{bm}
\usepackage{color}
\usepackage{amsmath}

\usepackage[breaklinks, hidelinks]{hyperref}
\usepackage{xcolor}
\hypersetup{
    colorlinks,
    linkcolor={red!50!black},
    citecolor={red!50!black},
    urlcolor={red!50!black}
}

\begin{document}
\title{Probing type-II Ising pairing using the spin-mixing parameter}

\author{Paulina Jureczko}
\affiliation{Institute of Physics, University of Silesia in Katowice, 41-500 Chorzów, Poland}
\author{Jozef Hani\v{s}}
\affiliation{Institute of Physics, Pavol Jozef \v{S}af\'{a}rik University in Ko\v{s}ice, Park Angelinum 9, 04001 Ko\v{s}ice, Slovakia}

\author{Paulo E. Faria~Junior}
\affiliation{Institute for Theoretical Physics, University of Regensburg, 93040 Regensburg, Germany}

\author{Martin Gmitra}
\email{martin.gmitra@upjs.sk}
\affiliation{Institute of Physics, Pavol Jozef \v{S}af\'{a}rik University in Ko\v{s}ice, Park Angelinum 9, 04001 Ko\v{s}ice, Slovakia}
\affiliation{Institute of Experimental Physics, Slovak Academy of Sciences, Watsonova 47, 04001 Ko\v{s}ice, Slovakia}

\author{Marcin Kurpas}
\email{marcin.kurpas@us.edu.pl}
\affiliation{
 Institute of Physics, University of Silesia in Katowice, 41-500 Chorzów, Poland 
}%

\date{\today}

\begin{abstract}

The immunity of Ising superconductors to external magnetic fields originates from a spin locking of the paired electrons to an intrinsic Zeeman-like field.
The spin-momentum locking in non-centrosymmetric crystalline materials leads to type-I Ising pairing in which the direction of the intrinsic field can be deduced from the spin expectation values. 
Conversely, in centrosymmetric crystals the electron spins locked to the orbitals can form Ising type-II pairs consisting of spin-orbit split doublets. Due to time-reversal symmetry, the doublets are spin degenerate, making it difficult to read the spin polarization of bands and the direction of spin-orbit fields. 
Here we present an efficient approach to determine the direction of the intrinsic field using the spin-mixing parameter $b^2$. 
Using first principles calculations based on the density functional theory, we study monolayer  
transition metal dichalcogenide superconductors PdTe$_2$, NbTe$_2$, and TiSe$_2$ with the 1T structure.
We calculate $b^2$ for individual Fermi pockets and provide a general picture of possible Ising type-II pairing within the full Brillouin zone. 
In order to complement our first principles results, we use group theory to provide a detailed picture of spin-orbit coupling and spin mixing in the relevant bands forming Fermi pockets.  
We demonstrate that contrary to the anticipated effects of spin-orbit locking, not every spin-orbit split spin doublet actively participates in Ising pairing.
Finally, by connecting the spin-mixing parameter $b^2$ with the intrinsic out-of-plane Zeeman field we estimate the upper in-plane critical magnetic field. 
\end{abstract}

\maketitle

\section{\label{sec:intro}Introduction} 

Quantum confinement in thin superconductors significantly eliminates orbital effects for an in-plane magnetic field leading to the enhancement of the upper critical fields \cite{tinkham_book}.
Superconductivity has recently moved from thin films to crystalline atomically thin systems \cite{Saito2016,Qiu_2021} conducting fascinating quantum phenomena. 
A recent breakthrough was the discovery of unconventional Ising pairing responsible for extremely large magnetic in-plane upper critical fields \cite{Li2021,Zhang2021}.
The Ising type-I 
superconducting pairing has been first identified in ionic-gated MoS$_2$ \cite{Lu2015,Saito2016} and further confirmed in NbSe$_2$ \cite{Xi2016,Xing2017} and in other trigonal prismatic polytypes transition metal dichalcogenides (TMDC) \cite{Moratalla2016,Tsen2016,Xing2017,DeLaBarrera2018,Lu2018,Tanaka2020}. 
An unconventional superconducting behavior has been also reported in 1T$^\prime$-MoS$_2$ \cite{Peng2019} and in few-layer stanene \cite{Liao2018} suggesting the Ising type-II pairing \cite{Falson2020}.

Spin-orbit coupling (SOC) combines orbital crystal symmetry and spin effects impacting the spin degeneracy of the electronic bands \cite{Dresselhaus1955,Elliott_1954}.  For crystals with broken inversion symmetry, SOC leads to spin splitting of the electronic bands except for the time-reversal invariant points,  while the band spin degeneracy is preserved in the centrosymmetric crystals. 
In the former case, the Fermi surface splits into two sheets with different densities of states at the Fermi level and independent superconducting pairing.
SOC can also lead to the mixing of singlet and triplet pairing  \cite{Gorkov2001,Frigeri2004} and to the enhancement of upper critical field in topological superconductors with space inversion symmetry \cite{Xie2020,Zhang2022}. It is thus an important ingredient for the emergence of unconventional electron pairing \cite{Sigrist_2005} in crystalline systems.

Based on the general band spin splitting, we distinguish two types of Ising superconductors in two-dimensional (2D) materials.
The Ising type-I superconductors \cite{Lu2015}, such as 1H polytype NbSe$_2$ or MoSe$_2$, have broken inversion symmetry but possess a horizontal mirror plane $\sigma_h$ which protects the spins in the perpendicular direction due to a strong SOC field $B_\text{so}$, also called an intrinsic Zeeman field. A giant in-plane upper critical magnetic field $B_{\parallel}$ is needed to compensate for the protecting $B_\text{so}$ \cite{Xi2016,Saito2016,DeLaBarrera2018}.
The robustness of the Ising pairing can be approximately estimated by comparing the spin splitting of the relevant bands close to the Fermi level, $2\Delta_{\text{so}}=\mu_B B_\text{so}$ with the Zeeman spin splitting $\Delta_{\rm Z}$ due to an external magnetic field $B_{\parallel}$.
The ratio $B_{\rm eff,\parallel}=B_{\parallel}^2/B_\text{so}$, defines an effective in-plane component of the magnetic field. When $B_{\rm eff,\parallel}$ reaches the Pauli paramagnetic limit  $B_{\rm eff,\parallel} \sim  B_{\text{P}} \approx 1.84\,T_{\rm c}$ \cite{Chandra1962,Clogston1962}, the Cooper pair breaking takes place destroying the superconducting condensate \cite{Xi2016}. 
From this condition, the upper critical field $B_{c2,\parallel}$ at $T=0$K can be estimated as  $B_{c2,\parallel} \approx \sqrt{ B_{\text{so}}B_{\text{P}}}$. 
For typical Zeeman splitting energy $\Delta_{\rm Z}\approx 0.1$\,meV/T (assuming $g_s \approx 2$), the values of $B_{c2,\parallel}$ in 2D TMDC superconductors can reach tens of Tesla, greatly exceeding $B_{\text{P}}$ \cite{Lu2015,Saito2016,Saito2016review,Xi2016,DeLaBarrera2018}.

Interestingly, Ising type-II superconductors can be realized in centrosymmetric 2D materials with multiple degenerate orbitals \cite{Wang2019}.
SOC field in those systems can induce spin-orbital locking contrary to the spin-momentum locking in type-I superconductors. But similarly, an induced intrinsic out-of-plane Zeeman-like field  $B_{\rm so}$ is insensitive to an in-plane magnetic field leading to a robust upper critical magnetic field \cite{Liu2020,Peng2019,Liao2018,Falson2020}. In centrosymmetric systems, for instance in 2D TMDCs with $D_{3d}$ point group symmetry \cite{Wang2019,Liu2020},  SOC splits all orbital states with orbital quantum number $\ell > 0$ and gives rise to spin degenerate doublets at the $\Gamma$ point. 
Energy splitting of the doublets is proportional to the effective Zeeman field $B_{\rm so} \tau_z$, where $\tau_z = \pm 1$ labels, e.~g., for $\ell=1$, $p_x \pm i p_y$ orbitals. The opposite signs of the field are related with the $p_x \pm i p_y$ orbitals. 
The presence of a $C_3$ rotation axis ensures the out-of-plane direction of the effective Zeeman field correspondingly polarizing the electron spins.
To determine the spin orientation within the spin degenerate bands one can break time-reversal symmetry by applying an effective Zeeman field, and compare its effect on band spin splitting without SOC, and with SOC split doublets \cite{Wang2019}.
This approach has been applied to identify a set of type-II Ising superconductor candidates investigating states close to the center of the Brillouin zone \cite{Wang2019}.

In this paper, we propose an alternative technique to determine the direction of the intrinsic Zeeman-like field in centrosymmetric systems which do not require breaking of inversion or time reversal symmetry. The proposed technique is fully ab initio and does not require auxiliary effective models.
The intrinsic Zeeman field direction is associated with the spin-mixing parameter $b^2$ reflecting spin-orbital locking and other features of SOC in a material.
By calculating the $b^2$ for different spin quantization axes we determine its anisotropy and conclude on the orientation of the intrinsic SOC field and spin polarization of degenerate bands. 

We test the proposed method using first principles calculations applied to three centrosymmetric 1T-polytype monolayer TMDCs classified as potential type-II Ising superconductors \cite{Wang2019}: PdTe$_2$, NbTe$_2$, TiSe$_2$. 
Our results confirm a strong Ising pairing in PdTe$_2$, with the in-plane upper critical fields $B_{\textrm{c2},\parallel}(T)$ up to 10 Tesla, in a perfect agreement with a recent experimental study \cite{Liu2020}.
Monolayer TiSe$_2$, besides significant out-of-plane spin polarization of bands around the Brillouin Zone center, is characterized by much weaker $B_{\textrm{c2},\parallel}(T)$, while for NbTe$_2$ the almost isotropic $b^2$ suggests no type-II Ising superconductivity. 

The proposed method is very versatile, can be easily implemented in numerical codes, and its application is not restricted only to 2D systems. It can be used to extend high-throughput screening calculations \cite{Wang2019} of 2D Ising type-II superconductors and can give an insight into SOC contribution to the critical field anisotropy in bulk superconductors.

\section{Methods}\label{sec:methods}
\subsection{Theoretical background }

Our approach is based on the fundamental property, that in centrosymmetric systems, the two spin degenerate Bloch states with crystal momentum $\mathbf{k}$,
\begin{equation}
\begin{array}{ll}
    \psi_{\mathbf{k},s}^{\Uparrow}(\mathbf{r}) &= \left[ a_{\mathbf{k},s}(\mathbf{r}) \chi^\uparrow_{s} + b_{\mathbf{k},s} (\mathbf{r})\chi^\downarrow_{s} \right ]\exp(i \mathbf{k}\cdot \mathbf{r}), \\
    \psi_{\mathbf{k},s}^{\Downarrow}(\mathbf{r}) &=\left[a^{*}_{-\mathbf{k}} (\mathbf{r})\chi^\downarrow_{s} -b^{*}_{-\mathbf{k}}(\mathbf{r}) \chi^\uparrow_{s}\right]\exp(i \mathbf{k}\cdot \mathbf{r}),
\end{array}
\end{equation}
are mixtures of spin \textit{up} and \textit{down} states,  $\chi^\uparrow_{s}$ and  $\chi^\downarrow_{s}$, respectively. Here, $a_{\mathbf{k},s}(\mathbf{r})$ and  $b_{\mathbf{k},s} (\mathbf{r})$ are lattice periodic functions, and $s$ stands for the spin quantization axis (SQA), $s=\lbrace x, y, z\rbrace$. 

The states  $\chi^\sigma_{s}$ are eigenstates of spin one-half operator $\hat{s}$, whose choice depends on the SQA. For instance, for SQA along the $x$ axis ($s=x$), $\chi^\sigma_{s}$ diagonalize the $\hat{s}_x$ Pauli spin matrix.

In the above Bloch states, the \textit{admixture coefficient} $b_{\mathbf{k},s}(\mathbf{r})$ is the amplitude of the spin component admixed by SOC to a pure spin state and vanishes if SOC is absent \cite{Elliott_1954}. The spin mixing parameter $b^{2}_{\mathbf{k},s}$ is defined as  the integral  of $b_{\mathbf{k},s}(\mathbf{r})$ over the whole unit cell
\begin{equation}
    b^{2}_{\mathbf{k},s} = \int_{V_{\rm cell}}|b_{\mathbf{k},s}(\mathbf{r})|^{2} d\mathbf{r}.
\end{equation}
For spin one-half particles, $b^{2}_{\mathbf{k},s}$ can be also expressed as a deviation of spin expectation value from its nominal value $1/2$ (in units of $\hbar$): 
\begin{equation}
    b^{2}_{\mathbf{k},s}=1/2 - |\langle \psi^{\sigma}_{\mathbf{k},s} | \hat{s} |\psi^{\sigma}_{\mathbf{k},s}\rangle |,
\end{equation} 
where $\sigma = \lbrace \Uparrow, \Downarrow\rbrace$  and $ \hat{s}$ is the spin one-half operator diagonal in the basis given by the choice of the SQA \cite{zimmermann2012}. For normalized states  $b^{2}_{\mathbf{k},s}$ varies from 0 for pure spinors to 0.5 for maximally spin mixed states.
Usually, it takes a small value due to a weakness of SOC but can be greatly enhanced at special points of the Brillouin zone, such as high symmetry points, accidental degeneracies or BZ edges, leading to a strong spin mixing \cite{fabian1998}. A more comprehensive discussion of the spin mixing parameter can be found in Refs. \cite{fabian1998,zimmermann2012,Mokrousov2013,Kurpas2019}.

The value of $b^{2}_{\mathbf{k},s}$ characterizes the strength of the spin-orbit interaction in a band, while its anisotropy for SQAs reflects the anisotropy of SOC and describes the preferred direction of the intrinsic SOC fields (Zeeman-like field) and spin polarization \cite{zimmermann2012}. Thus, calculating $b^{2}_{\mathbf{k},s}$ and its anisotropy for individual Fermi pockets should infer their potential contribution to the Ising pairing.

\subsection{Details of first principles calculations}

The monolayers of the 1T-polytypes of TMDC are derived from the bulk variants crystallizing in space group 164 with metal atoms (Pd, Nb, Ti) in Wyckoff position 1a (0,0,0), and the chalcogen atoms (Se,Te) in 2d positions (2/3, 1/3, $d$). We considered vacuum spacing of 20~{\rm\AA} to eliminate periodic images in the vertical direction.

The first principles calculations were performed using the open source code suite {\sc Quantum Espresso} \cite{QE-2009,Giannozzi_2017}, implementing pseudopotential and plane wave approaches to the density functional theory. We used scalar and fully relativistic SG15 Optimized Norm-Conserving Vanderbilt (ONCV) pseudopotentials \cite{Hamman2017,Scherpelz2016}. The kinetic energy cutoffs for the wave function were 52\,Ry, 53\,Ry, 52\,Ry for PdTe$_2$, NbTe$_2$, and TiSe$_2$ respectively, while cutoffs for the charge density were four times bigger. These values were sufficient to obtain well-converged values of orbital and spin-related quantities. 

The lattice parameters and atomic positions in the unit cell were optimized to minimize strain and internal forces. The lattice constants were determined by finding the minimum of a quadratic function fitted to the first principles data of total energy calculated for several values of $a$. 
Later, the positions of atoms were fully relaxed using the quasi-Newton scheme as implemented in {\sc Quantum Espresso} package. The optimized lattice constants and vertical positions of the chalcogen atoms are $a=4.02\,{\rm\AA}$, $d=1.4\,{\rm\AA}$ for PdTe$_2$, $a=3.67\,{\rm\AA}$, $d=1.86\,{\rm\AA}$ for NbTe$_2$, and $a=3.54\,{\rm\AA}$, $d=1.55\,{\rm\AA}$ for TiSe$_2$,
and agrees well with the experimental values; for PdTe$_2$, TiSe$_2$ in monolayer \cite{Liu2020, Peng_15} and for NbTe$_2$ in ultrathin nanoplates \cite{Li2018}.
Self-consistency was obtained with $21\times21\times1$ Monkhorst-Pack grid sampling of the first Brillouin zone (BZ).
The Fermi contour averages of the spin-mixing parameter $b^2$ were calculated using the tetrahedron method with an adaptive mesh of $k$-points according to the formula
\begin{equation}
\label{av_val}
b^2_s \equiv \langle b^2_{\mathbf{k},s}\rangle_{\rm FC}  = \dfrac{1}{\rho(E_{F})S_{\rm BZ}} \int_{\rm FC} \dfrac{b^2_{\mathbf{k},s}}  {\hbar |\boldsymbol{v}_{F}(\mathbf{k})|}d^2k,
\end{equation}
where $S_{\rm BZ}$ is the area of the Fermi surface within the first Brillouin zone,
FC stands for Fermi contour, 
$\rho(E_{F})$ is the density of states per spin at the Fermi level, and  $\boldsymbol{v}_{F}(k)$ is the Fermi velocity.
The mesh density was iteratively increased, using as a criterion variation of $b^2_{\mathbf{k}}$, until convergence of the average $b^2$ has been reached. The number of generated $k$-points in each Fermi pocket varied between 2000 and 5000. 

  \section{Results}\label{sec:results}

\subsection{First principles calculations}

\begin{figure*}[t]
    \centering
    \includegraphics[width=0.99\textwidth]{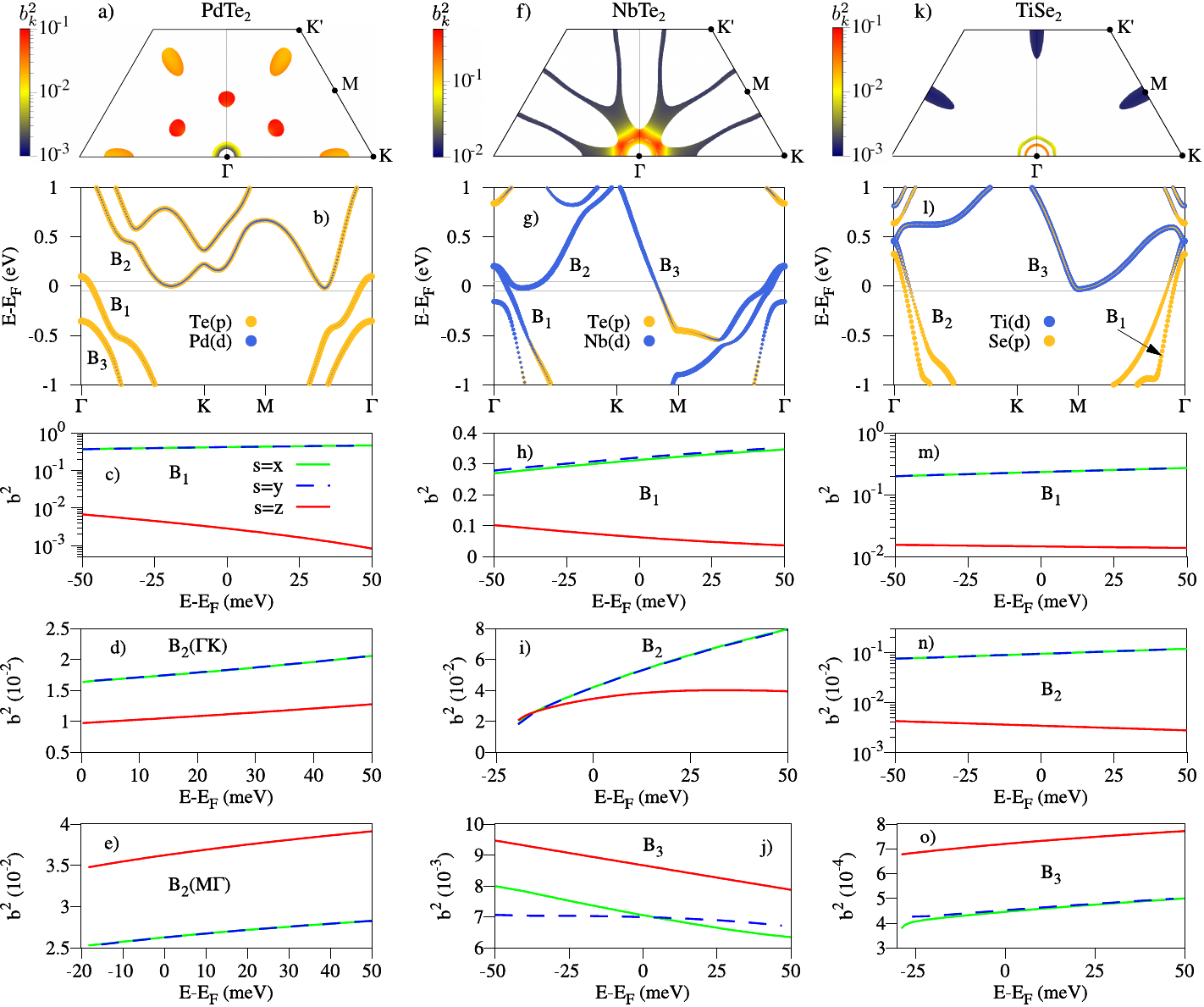}
    \caption{Calculated orbital and spin-orbital properties of 1T-polytype monolayers of transition metal dichalcogenides.
    (a)~band pockets in energy window $\pm 50$ meV around the Fermi energy for PdTe$_2$ in the Brillouin zone wedge. The colormap corresponds to $b^2_\mathbf{k}$ for SQA $s=z$.
    (b)~orbital resolved band structures of PdTe$_2$ with the indicated bands crossing the Fermi energy. Grey horizontal lines depict the $\pm 50$ meV energy window around the Fermi energy for which the average spin-mixing parameter $b^2$ was calculated.
    (c), (d), (e)~Fermi contour averaged spin-mixing parameter $b^2$ calculated for the Fermi pockets formed by the bands $B_1$ and $B_2$ depicted in (a) for three spin quantization axes $s=\lbrace x,y,z\rbrace$ aligned with the Cartesian axes. Similarly, the middle and right row for NbTe$_2$ and TiSe$_2$.
    }
    \label{fig:bs_b2}
\end{figure*}

In this section, we present the results of our first principles calculations. First, we discuss PdTe$_2$. Among all materials studied here, it is the only one for which the in-plane upper critical field has been measured \cite{Liu2020}. Therefore it represents a platform to test the proposed approach.
We mention that PdTe$_2$ crystal is topologically non-trivial material and host Dirac type-II Fermions \cite{Noh2017,Fei2017}. It  becomes a superconductor around $T_{\rm c} = 1.7$~K \cite{Guggenheim1961}. 
Recent measurements on a few-layer PdTe$_2$ sample have shown, that the Pauli limit for the critical magnetic field is of $B_{\rm P}\approx 1.29$\,T for 4-layer and $B_{\rm P}\approx 1.36$\,T for 6-layer thick samples exceeded 7 times the Pauli limit corroborating the Ising type-II pairing \cite{Liu2020}.

In Fig.~\ref{fig:bs_b2}(b) we show the calculated orbital resolved relativistic band structure of monolayer PdTe$_2$. In this energy range, the bands of PdTe$_2$ are formed mainly by the $p$-electrons of tellurium with a small admixture of Pd $d$-electrons the band $B_2$ and bands of higher energy.  In the vicinity of the $\Gamma$ point, the band $B_1$ that crosses the Fermi energy is a linear combination of $p_x$ and $p_y$ Te electrons, up to the anticrossing at the energy around $-0.5$\,eV, when the contribution from $p_z$ electrons starts to be visible (see also Fig. S2 in the Supplemental Material \cite{supp}).

Recent explanation of the Ising type-II pairing \cite{Falson2020,Wang2019,Wang2021} involves the spin-orbit split-off orbital doublet $(p_x \pm ip_y,\sigma)$ around the $\Gamma$ point.
For PdTe$_2$ 

the orbital doublet is formed by the band $B_1$ with the total angular momentum $J_z=3/2$ and the band $B_3$ with  $J_z=1/2$, which without SOC are four-fold degenerate at the $\Gamma$ point see Fig.~S1(a) in the Supplemental Material \cite{supp}. The split-off energy is proportional to the SOC strength and is a direct measure of the spin-orbit locking of the spins in out-of-plane direction \cite{Liu2020}.    
The band resolved calculation of the $b^2$ for the $B_1$ band, see Fig.~\ref{fig:bs_b2}(c), reveals a suppression of the $b^2$ parameter for the SQA along the $z$ direction in comparison to the in-plane directions. 
For the latter, $b^2 \approx 0.5$, which happens when 
$\vert a_{\mathbf{k},x/y}(\mathbf{r})\vert \approx \vert b_{\mathbf{k},x/y} (\mathbf{r})\vert $, and  the spin expectation values for the in-plane components $|\langle \psi^{\sigma}_{\mathbf{k},x/y} | \hat{s}_{x/y} |\psi^{\sigma}_{\mathbf{k},x/y}\rangle| \approx 0 $ \cite{zimmermann2012}, which clearly demonstrates the out-of-plane polarization of spins and the intrinsic SOC field. 

It is convenient to express the anisotropy of the spin mixing parameter $b^2$ by the polarization $P$ in the following form 
\begin{equation}
P=\dfrac{b^2_\parallel - b^2_\perp}{b^2_\parallel + b^2_\perp},
\end{equation}
where $b^2_\parallel = (b^2_x + b^2_y)/2$ and $b^2_\perp = b^2_z$. The polarization is valued $-1\leq P \leq 1$ from in-plane to out-of-plane direction, while for $P=0$ no anisotropy is expected. From the above definition, it is clear, that Ising pairing is effective only when $P$ is close to one. 

In Fig.~\ref{fig:pdte_4plot_aver}(b) we show $P$ calculated for the $B_1$ band of PdTe$_2$ versus doping up to the edge of this band at the $\Gamma$ point. The strong out-of-plane spin polarization, $P\approx 1$, supports the observed Ising type-II pairing in this case \cite{Liu2020}.

The band $B_2$ forms two pockets along $\Gamma K$ and $\Gamma M$ lines with a mixture of all Te $p$ electrons and a small admixture of Pd $d$-electrons at the vicinity of the Fermi level (Fig.~\ref{fig:bs_b2}(b) and Fig.~S2 in the Supplemental Material \cite{supp}). The calculated $b^2$ for both the pockets shows almost equal amplitudes for all the SQA components, see Fig.~\ref{fig:bs_b2}(d) and Fig.~\ref{fig:bs_b2}(e), which leads to small anisotropy of $b^2$. The polarization $P$ for the pocket along the $\Gamma K$ line is $P\approx 0.23$ and for the pocket along the $\Gamma M$ line $P\approx -0.17$, which indicates that Ising pairing in the band $B_2$ is ineffective. This result is consistent with the common interpretation that type-II Ising superconductivity requires multiple degenerate bands.

The large polarization for the $B_1$ band suggests that a hole doping of the PdTe$_2$ should eliminate the $B_2$ pockets and provide enhancement of the Ising type-II pairing due to large $b^2$ anisotropy.
For heavy hole doping due to the presence of a substrate the $B_3$ band being the orbital doublet partner of the band $B_1$, may potentially give rise to the Ising pairing. The calculated spin-mixing parameter $b^2$  and polarization $P$ show, however, that the Ising mechanism is not present in the band $B_3$ due to the dominant in-plane direction of spin-orbit fields and weak anisotropy, see Fig.~S3 in the Supplemental Material \cite{supp}. We explain the weak anisotropy in this band in the Sec.~\ref{sec:mixing_theory}.

Superconductivity in niobium dichalcogenides down to monolayer limit has been extensively studied \cite{Frindt1972,Nagata1993,Xi2016,Wang2017,DeLaBarrera2018,Yan_2019}.
NbTe$_2$ is a polytypic TMDC exhibiting peculiar 1T distortions and pressure-induced phase transitions \cite{S.Li2021}. Recently cathodic exfoliation of NbTe$_2$ shows a trigonal structure of the flakes \cite{Li2021NM}. NbTe$_2$ belongs to the group 5 TMDCs nominally having one $d$-electron and is metallic \cite{Lasek2021}. The superconducting temperature was reported at $T_c = 0.5 – 0.72$~K \cite{VanMaaren1967, Nagata1993, Zhang2019}.
In Fig.~\ref{fig:bs_b2}(f) we plot $b^2_\mathbf{k}$ in the Brillouin zone wedge within the pockets for energies $\pm 50$~meV around the Fermi level. The corresponding averaged $b^2$ dependencies for the indicated bands in Fig.~\ref{fig:bs_b2}(g) are plotted in Fig.~\ref{fig:bs_b2}(h-j). The Bloch states near the Fermi level are predominantly formed by the $d$-orbitals of Nb atoms. 
The band resolved averaged $b^2$ indicates a rather small anisotropy, thus leading to reduced polarization values, $|P|\in [0,0.75]$. 

Due to this small anisotropy of $b^2$, we speculate that the NbTe$_2$ is not a good candidate to be a type-II Ising superconductor.

Two-dimensional TiSe$_2$ has been extensively studied to probe
the interplay between superconductivity and possible charge-density waves via doping \cite{Hu2021,Luo2015,May2011,Giang2010}, intercalation \cite{Adam2022,Piatti2023,Lee2021,Morosan2006,Morosan2007}, pressure \cite{Kusmartseva2009,Friend1982}, and layer thickness \cite{Li2016,Chen2015}. The charge-density wave phase can be suppressed by intercalation \cite{May2011,Morosan2006} or pressure \cite{Kusmartseva2009}, and superconducting critical temperature ranges $T_c = 0.7 - 3.9$~K \cite{Lee2021}.
The calculated $b^2_\mathbf{k}$ is shown in Fig.~\ref{fig:bs_b2}(k) for the Fermi pockets within the energy window $\pm 50$ meV formed by the three bands, $B_1$ and $B_2$ around the $\Gamma$ point, and $B_3$ around the $M$ point. 
The orbital resolved band structure of TiSe$_2$ shown in Fig.~\ref{fig:bs_b2}(l) reveals that the states near the $M$ points are formed by the Ti $d$-orbitals. The remaining bands are formed by the $p$-orbitals of Se atoms, and $d$-orbitals of Ti in the case of the band $B_2$, see Fig.~S4 in the Supplemental Material \cite{supp}.
The band resolved spin-mixing parameters for the $B_1$ and $B_2$ bands, see Fig.~\ref{fig:bs_b2}(m,n) exhibit distinct but smaller anisotropy than $B_1$ of PdTe$_2$,  within the whole doping range,  $b^2_{\parallel}/b^2_{\perp}\approx 10-30$. The values of $P$ vary between 0.85 and 0.9 for the $B_1$ band and between 0.9 and 0.95 for the $B_2$ band, see Fig.~S6 in the Supplemental Material \cite{supp}. 
In comparison, for the $B_3$ band $b^2_{\parallel}/b^2_{\perp}\approx 2$, see Fig.~\ref{fig:bs_b2}(o).
Thus, one can expect that only bands $B_1$ and $B_2$ will give a relevant contribution to the Ising enhancement of the upper critical fields in TiSe$_2$.
%
%
\subsection{The origin of spin mixing}\label{sec:mixing_theory}
To understand the mechanism of spin mixing in the studied materials, we analyze the spin-orbit coupling in the band structure using symmetry arguments, following Refs.~\cite{Kurpas2019, FariaJunior2022NJP}. We discuss in detail PdTe$_2$ here in the main text, and the analogous analysis for NbTe$_2$ and TiSe$_2$ can be found in the Supplemental Material \cite{supp}.

The bands relevant for Ising pairing are $B_1$ and $B_3$ close to the $\Gamma$ point. Without SOC these bands are degenerate and transform as the $\Gamma_3^-$ irrep of the $D_{3d}$ point group, see Fig.~S1(a) in the Supplemental Material \cite{supp}. 
The intraband SOC leads to the Hamiltonian
\begin{equation}\label{eq:Hintra}
    H_\text{intra} = -\Delta_z s_z \sigma_y,
\end{equation} 
 in the basis 
 $\lbrace |\Gamma_{3,1}^{-} \uparrow\rangle, |\Gamma_{3,2}^{-} \uparrow\rangle, |\Gamma_{3,1}^{-} \downarrow \rangle, |\Gamma_{3,2}^{-} \downarrow\rangle\rbrace$
with $\Gamma_{3,1}^{-} \sim x$ and $\Gamma_{3,2}^{-} \sim y$, 
 $\Delta_z$ is the SOC strength, $s_z$ is spin one-half operator and $\sigma_y$ is the Pauli matrix in $y$ direction acting in the orbital subspace. The Hamiltonian (\ref{eq:Hintra}) splits off the bands by $2 |\Delta_z|$, but leaves the spin unchanged, giving no spin mixing ($b^2=0$). The eigenstates of $H_\text{intra}$ are pure $s_z$ spinors,  $\left|\Gamma_{3\pm}^{-}\sigma\right\rangle   =\frac{1}{\sqrt{2}}\left(\left|\Gamma_{3,1}^{-}\right\rangle \pm i\left|\Gamma_{3,2}^{-}\right\rangle \right)\left|\sigma\right\rangle $, $\sigma=\lbrace\uparrow,\downarrow\rbrace$, due to time-reversal symmetry which does not allow spin-flip terms in (\ref{eq:Hintra}). 
Mixing of spins originates from the interband SOC, for which no such restriction applies. Treating SOC perturbatively (see the Supplemental Material \cite{supp} for details) we get new wave functions of the SOC split doublets $B_1$ and $B_3$
\begin{equation}
\label{eq:eigenstates_so}
\begin{array}{ll}
    \left| \Gamma_{3+}^{-}\Uparrow\right\rangle_{B_1} &=\left|\Gamma_{3+}^{-}\uparrow\right\rangle +\alpha\left|\beta\uparrow\right\rangle   + \gamma \left|\beta\downarrow\right\rangle ,  \\
     \left|\Gamma_{3-}^{-}\Downarrow\right\rangle_{B_1} &=\left|\Gamma_{3-}^{-}\downarrow\right\rangle +\alpha \left|\beta\downarrow\right\rangle   + \gamma \left|\beta\uparrow\right\rangle , \\
      \left|\Gamma_{3+}^{-}\Downarrow\right\rangle_{B_3} &=\left|\Gamma_{3+}^{-}\downarrow\right\rangle +\delta\left|\beta\downarrow\right\rangle   + \zeta \left|\beta\uparrow\right\rangle  ,\\
      \left| \Gamma_{3-}^{-}\Uparrow\right\rangle_{B_3} &=\left|\Gamma_{3-}^{-}\uparrow\right\rangle +\delta\left|\beta\uparrow\right\rangle   + \zeta \left|\beta\downarrow\right\rangle,   
    
\end{array}
\end{equation}
where $\alpha, \gamma, \delta,\zeta$ are complex coefficients. It is clear, that the states (\ref{eq:eigenstates_so}) are mixtures of spin \textit{up} and spin \textit{down} components giving $b^2\neq 0$.  
Different coefficients in the states (\ref{eq:eigenstates_so}) indicate different spin mixing in the bands $B_1$ and $B_3$. Indeed, our theoretical analysis reveals that spin mixing in the band $B_1$  results from the coupling of the $\Gamma_3^-$ band to other $\Gamma_3^-$ bands, while in $B_3$ it comes from the coupling to the bands with $\Gamma_1^{-}$ and $\Gamma_2^-$ irreducible representations. 
In PdTe$_2$ the closest  $\Gamma_3^-$ band lies 6.5~eV above the relevant $\Gamma_3^-$ band, and the intraband SOC dominates over the weak interband coupling, which explains the small value of $b^2_z$ and large anisotropy of spin mixing. The weak anisotropy of $b^2$ in the band $B_3$ is a common effect of the intraband spin-conserving SOC and the interband spin-flip coupling between the $\Gamma_3^-$ and the nearby $\Gamma_2^-$ band (see Fig.~\ref{fig:tc2_maps}(a) in the Supplemental Material \cite{supp}).

According to the recent interpretation of the Ising pairing mechanism involving  degenerate orbitals, due to spin-orbit locking  the spin-orbit split doublets, such as $B_1$ and $B_3$ in PdTe$_2$, 
should display a perfect out-of-plane spin polarization \cite{Liu2020,Falson2020}. Our numerical results for $b^2$ supported by symmetry analysis show, that such a picture is simplified. The doublet partner bands exhibit considerably different spin polarization direction due to coupling to different bands via SOC. The spin mixing parameter $b^2$ detects these differences, which proves the reliability of the proposed method in detecting Fermi contours relevant for possible Ising pairing. 

\subsection{Estimation of the upper critical field $B_{c2,\parallel}$}
To stress the experimental relevance of our method we will now estimate the upper critical field $B_{c2}$ for PdTe$_2$. Monolayer PdTe$_2$ crystallizes in $P3\bar{m}1$ space group being isomorphic with $D_{3d}$ point group.  At the Brillouin zone center, the presence of the $C_{3}$ rotation axis forbids the in-plane spin components and ensures the out-of-plane orientation of the intrinsic SOC field. 
Away from the $\Gamma$ point, such restriction is released and all components of the intrinsic Zeeman field $\mathbf{B}$ are allowed, $\mathbf{B}=(B_x,B_y, B_z)$.
Next, we make an assumption, that the main contribution to the splitting of the bands $B_1$ and $B_3$ comes from the intraband SOC, while the interband SOC Hamiltonian contributes mainly to spin expectation values and spin mixing. This assumption is reliable as long as $b_z^2 \approx 0 \Rightarrow P \approx 1$, which also allows us to select the relevant bands that are more susceptible to host Ising paring superconductivity. Furthermore, for bands with $b_z^2 \approx 0 \Rightarrow P \approx 1$, a g-factor of $\approx 2$ is expected, since they would share nearly the same orbital character with opposite spins \cite{Raiber2022NatComm}.
With this picture in mind, the splitting energy can be written as $\Delta_{\rm{so}}=2|\Delta_z|=g_s \mu_B B$, where $B =\|\mathbf{B}\|= \sqrt{B_x^2 + B_y^2 + B_z^2}$, and $g_s=2$, is the electron Lande factor. 
We are interested in the in-plane to out-of-plane anisotropy, thus it is convenient to rewrite $\Delta_{\text{so}}$ as $\Delta_{\text{so}}=2\mu_B \sqrt{B_\parallel^2 + B_\perp^2}$, where  $B_{||} = \sqrt{B_x^2 +B_{y}^2 }$ and $B_\perp=B_z$. 
We can now relate the direction of  $\mathbf{B}$ to the polarization vector $P$
\begin{eqnarray}\label{eq:defB}
    B_\parallel &=& \frac{B}{\sqrt{2}\sqrt{1+P^2}} \left(1-P\right),\\
    B_\perp &=& \frac{B}{\sqrt{2}\sqrt{1+P^2}} \left(1+P\right).
\end{eqnarray}
Since the Ising pairing is most effective when $B_\parallel\approx 0$ and $B_\perp \approx 1$ (or $P\approx 1$) we can consider only $B_\perp$. Replacing  $B$  in (\ref{eq:defB}) with $\Delta_{\rm so}/2\mu_B$ gives
\begin{equation}\label{eq:B_perp}
     B_\perp =\frac{\Delta_{\rm so}}{2\mu_B} \frac{1}{\sqrt{2}\sqrt{1+P^2}} \left(1+P\right),
\end{equation}
where $P$ and  $\Delta_{\rm so}$  are extracted from our first principles calculations. Away from from the $\Gamma$ point $\Delta_{\rm so}$ at a given $\mathbf{k}$-point  is calculated using the formula
\begin{equation}
    \Delta_{\rm so}  = \Delta_{n,m}^{\rm rela} - \Delta_{n,m}^{\rm nrel},
\end{equation}
where $\Delta_{n,m}^{\rm rela/nrel}=E_m - E_n$ is the energy distance between the bands $n$ and $m$ forming the spin-orbit split doublets, and superscript refers to relativistic/non-relativistic calculations, i. e., calculations with and without SOC, respectively. For PdTe$_2$ the indices $m$ and $n$ correspond to the $B_1$ and $B_3$ bands, respectively.

Knowing $B_\perp$ we can estimate the in-plane upper critical field  $B_{\textrm{c,2},\parallel}$ at 0 K following Ref. \cite{Xi2016}
\begin{equation}\label{eq:bc2_0K}
    B_{\textrm{c2},\parallel} \sim \sqrt{ B_{\rm P}B_{\perp}}.
\end{equation}

\begin{figure}[h]
    \centering
    \includegraphics[width=0.99\columnwidth]{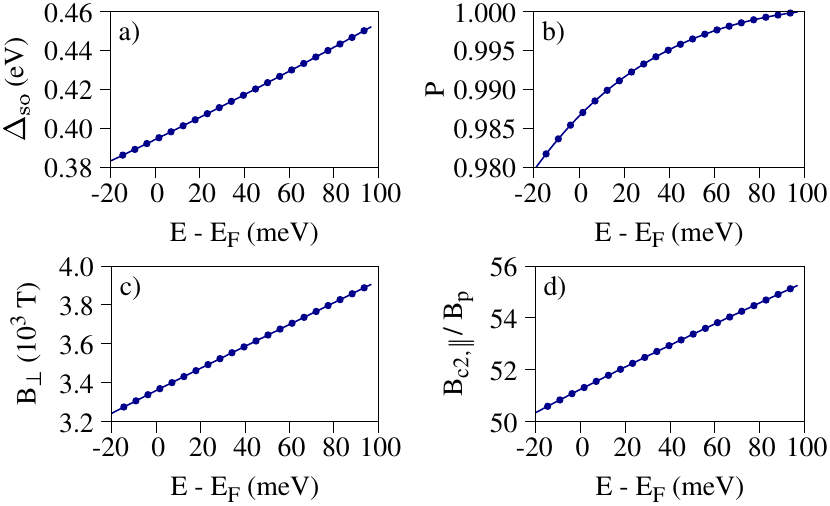}
    \caption{Calculated properties for monolayer PdTe$_2$ at 0~K as a function of the Fermi energy. 
    (a)~The Fermi contour averaged spin-orbital splitting $\Delta_{\rm so}$,
    (b)~polarization vector $P$, 
    (c)~intrinsic out-of-plane Zeeman field $B_\perp$ and 
    (d)~in-plane upper critical field $B_{c2,\parallel}$ in units of Pauli field $B_{\rm P}$.} 
    \label{fig:pdte_4plot_aver}
\end{figure}

The results are shown in Fig. \ref{fig:pdte_4plot_aver}. Due to the large $\Delta_{\rm so}$ and $P\approx 1$,  the values of $B_\perp$ are giant, on the order of $10^3$~Tesla, 
resulting in $B_{\textrm{c2},\parallel}$(0\;K) exceeding the Pauli limit by more than fifty times. The large value of $B_\perp$ is mainly determined by large $\Delta_{\text{so}}$, since in the whole range of doping $P\approx 1$. 

To estimate $B_{\textrm{c2},\parallel}$ at finite temperatures we use the microscopic model for the one-band Ising superconductivity from Ref. \cite{LiuPRX_2018}, which was used in Ref. \cite{Liu2020} to extract $B_{\textrm{c2},\parallel}$ for PdTe$_2$. Using the same model and material parameters (see Supplemental Material \cite{supp} for details) will give us a direct comparis on of our method with the experimental data.
 
\begin{figure}[h]
    \centering
    \includegraphics[width=0.85\columnwidth]{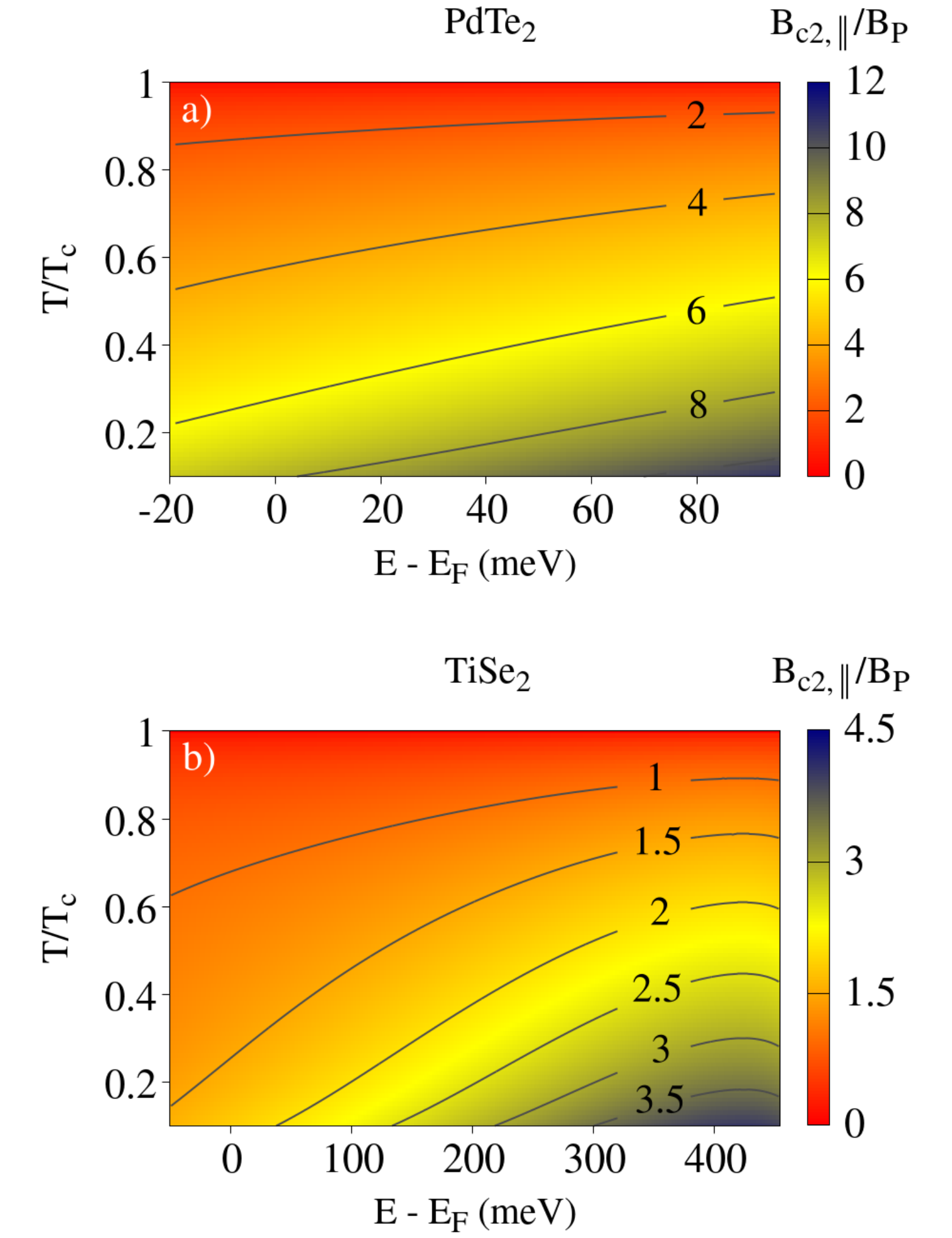}
    \caption{The upper critical field $B_{c2,||}$  versus temperature and the Fermi level for (a) PdTe$_2$ and (b) TiSe$_2$ for band B$_2$ estimated using Eq. (14) from the Supplemental Material \cite{supp}.}
    \label{fig:tc2_maps}
\end{figure}

The calculated values of the in-plane upper critical field $B_{\textrm{c2},\parallel}(T)$, see Fig. \ref{fig:tc2_maps} (a), are in very good agreement with the experimental values, taking into account small differences between the number of layers and other factors present in the experimental setup. 
For 6-layer PdTe$_2$ the experimental values of $B_{\textrm{c2},\parallel}(T)$ vary from 6\,$ B_{\textrm{P}}$ to 8\,$B_{\textrm{P}}$ for $T\in$[0.2\;K,0.6\;K] \cite{Liu2020}. Our theoretical values for the same temperatures are slightly smaller due to momentum dependence of the spin-orbital gap $\Delta_{\textrm{so}}$. Taking $\Delta_{\textrm{so}}(\Gamma)=0.45$\,eV we get  $B_{c2,\parallel}(T)$ between 5\,$B_{\textrm{P}}$ and 10\,$B_{\textrm{P}}$, pointing for stronger out-of-plane spin polarization of bands.  We note, that we did not use here any free or fitting parameter. 

Similar analysis performed for the bands $B_1$ and $B_2$ of TiSe$_2$ indicates that despite relatively large $\Delta_{\textrm{so}}$ the values of $B_{c2,\parallel}$(0\;K) and $B_{c2,\parallel}(T)$ are considerably smaller than for PdTe$_2$, see (Fig.~ 3 (b) and Fig.~S6 and Fig.~S7 in the Supplemental Material \cite{supp}, indicating that the Ising pairing in TiSe$_2$ would require relatively low temperatures and high electron doping.

\section{Conclusions}
\label{sec:conclusions}

We have proposed an efficient approach to analyze the direction of intrinsic spin-orbit coupling fields and spin polarization in the band structure using the anisotropy of the spin-mixing parameter $b^2$. This information is crucial for understanding the impact of type-II Ising pairing in superconductors with space inversion centers.
By applying first principles calculations, we tested this approach on three monolayer transition metal dichalcogenide superconductors in the T phase: PdTe$_2$, NbTe$_2$, and TiSe$_2$. 
The largest anisotropy of $b^2$, and the strongest expected Ising type-II pairing, is found for PdTe$_2$, for which our results of $B_{\textrm{c2},\parallel}$ are in excellent agreement with experimental values.
Furthermore, TiSe$_2$ is predicted to be also a good candidate for type-II Ising superconductivity but requires considerate electron doping. In contrast to PdTe$_2$ and TiSe$_2$, we found that NbTe$_2$  shows a moderate anisotropy of $b^2$, insufficient to observe Ising pairing in this material.

The main advantage of the present approach is that it can detect possible contributions to the Ising superconductivity from individual Fermi pockets irrespective of their orbital composition and position in the Brillouin zone. As $b^2$ is calculated directly from the wave functions and contains all symmetry properties of the crystal, the method is very versatile and independent of the basis set.
Even though we can not provide an answer to superconductivity formation in the individual Fermi pockets, by inspecting the anisotropy of $b^2$ one can easily select potentially relevant $k$ momenta for type-II Ising superconductors.

\begin{acknowledgments}
P.J. and M.K. acknowledge support from the Interdisciplinary Centre for Mathematical and Computational Modelling (ICM), University of Warsaw (UW), within grant no. G83-27. The project is co-financed by the Polish National Agency for Academic Exchange and by the National Center for Research and Development (NCBR) under the V4-Japan project BGapEng. J.H. and M.G. acknowledge financial support from the Slovak Research and Development Agency provided under Contract No. APVV-SK-PL-21-0055 and by the Ministry of Education, Science, Research and Sport of the Slovak Republic provided under Grant No. VEGA 1/0105/20 and Slovak Academy of Sciences project IMPULZ IM-2021-42 and project FLAG ERA JTC 2021 2DSOTECH. P.E.F.J. acknowledges the financial support of the Deutsche Forschungsgemeinschaft (DFG, German Research Foundation) SFB 1277 (Project-ID 314695032, projects B07 and B11) and SPP 2244 (Project-ID 443416183).

\end{acknowledgments}
\bibliography{bibliography}

\end{document}